\documentclass[12pt]{article}
\usepackage{graphicx} 
\usepackage{amsmath}
\usepackage{amssymb}
\usepackage{cite}
\usepackage{geometry}
\usepackage{booktabs}
\usepackage{array}
\usepackage{multirow}
\usepackage{paracol}
\usepackage{longtable}
\usepackage{hyperref}
\usepackage{caption}
\usepackage{graphicx}
\usepackage{float}
\usepackage{tikz}
\usetikzlibrary{shapes.geometric, arrows}
\usepackage{booktabs}  

\title{\textbf{Deep Learning-Based Automated Workflow for Accurate Segmentation and Measurement of Abdominal Organs in CT Scans}}

\author{%
    Dr.Praveen Shastry, Dr.Ashok Sharma, Kavya Mohan, Naveen Kumarasami, \\
    Anandakumar D, Mounigasri M, Keerthana R, Kishore Prasath Venkatesh, \\
    Bargava Subramanian, Kalyan Sivasailam
}

\date{}

\usepackage{ragged2e}  
\usepackage{titlesec}  

\titleformat{\section}{\raggedright\Large\bfseries}{}{0em}{}
\titleformat{\subsection}{\raggedright\large\bfseries}{}{0em}{}

\begin{document}

\maketitle
\section{Abstract}

\textbf{Background}: Automated analysis of CT scans for abdominal organ measurement is crucial for improving diagnostic efficiency and reducing inter-observer variability. Manual segmentation and measurement of organs such as the kidneys, liver, spleen, and prostate are time-consuming and subject to inconsistency, underscoring the need for automated approaches.
\textbf{Purpose}: The purpose of this study is to develop and validate an automated workflow for the segmentation and measurement of abdominal organs in CT scans using advanced deep learning models, in order to improve accuracy, reliability and efficiency in clinical evaluations.
\textbf{Methods}: The proposed workflow combines nnU-Net, U-Net++ for organ segmentation, followed by a 3D RCNN model for measuring organ volumes and dimensions. The models were trained and evaluated on CT datasets with metrics such as precision, recall and Mean Squared Error (MSE) to assess performance. Segmentation quality was verified for its adaptability to variations in patient anatomy and scanner settings.\\
\textbf{Results}: The developed workflow achieved high precision and recall values, exceeding 95 for all targeted organs. The Mean Squared Error (MSE) values were low, indicating a high level of consistency between predicted and ground truth measurements. The segmentation and measurement pipeline demonstrated robust performance, providing accurate delineation and quantification of the kidneys, liver, spleen, and prostate.\\
\textbf{Conclusion}: The proposed approach offers an automated, efficient, and reliable solution for abdominal organ measurement in CT scans. By significantly reducing manual intervention, this workflow enhances measurement accuracy and consistency, with potential for widespread clinical implementation. Future work will focus on expanding the approach to other organs and addressing complex pathological cases.

\section{Introduction}
Automated analysis of medical imaging has become a pivotal area in the healthcare domain, offering significant improvements in diagnostic efficiency and accuracy [1].Computed Tomography (CT) is a widely used imaging modality for evaluating abdominal organs, which plays a crucial role in diagnosing and monitoring various health conditions, such as liver disease, kidney dysfunction, and prostate enlargement [2].Manual segmentation and measurement of these organs are time-consuming and prone to variability between observers, highlighting the need for automated approaches [3].
In this study, we present an automated workflow for the segmentation and measurement of the kidneys, liver, spleen, and prostate in CT scans using advanced deep-learning models [4]. The proposed pipeline combines state-ofthe-art architectures like nnU-Net, U-Net++ for segmentation, followed by 3D RCNN for organ measurement[5]. The goal is to develop a reliable, efficient solution that can assist radiologists in providing accurate and consistent measurements, ultimately improving clinical decision-making and patient outcomes
[6].
The use of nnU-Net and its complementary architectures allows an effective handling of variability in CT datasets, including differences in resolution, scanner settings, and patient anatomy [7]. The 3D RCNN model complements this by providing accurate quantification of organ parameters, which is critical foDou et al.2017r diagnosing abnormalities and tracking disease progression[8]. Evaluation metrics, including precision, recall and Mean Squared Error (MSE), are used to validate the performance of the workflow, with results demonstrating high accuracy and reliability [9].
The contributions of this research include the development of an end-to-end automated system for abdominal organ measurement, a detailed evaluation of the performance of advanced segmentation and measurement models, and the demonstration of the system’s potential for clinical implementation [10]. This paper aims to advance the state-of-the-art in medical image analysis, providing a solution that minimizes manual effort while delivering reliable and consistent results.

\section*{ Methodology :}

\subsection*{AI System Overview}

The AI system developed for abdominal organ measurement in CT scans is designed to be
an automated, efficient, and reliable solution for clinical implementation \cite{Gibson et al.2018}.
The system comprises two primary phases: segmentation and measurement.

\begin {itemize}

    \item \textbf{Segmentation Phase:}The segmentation phase utilizes deep learning architectures
like nnU-Net, U-Net++, and Nested U-Net to accurately segment organs such as the
liver, kidneys, spleen, and prostate \cite{Roth et al.2012}. These models are tailored to
manage variations in patient anatomy, image resolution, and scanner differences,
ensuring the robustness of segmentation outputs \cite{Roth et al.2016}. Each of these
models employs encoder-decoder structures, multi-scale feature extraction, and skip
connections to enhance segmentation accuracy \cite{Ren et al.2017}. The models were
trained on a diverse dataset to maximize generalizability and adaptability \cite{Chen et al.2018}.

\end{itemize}

\begin{itemize}
    \item \textbf{Measurement Phase: }Following segmentation, the measurement phase employs a
3D RCNN model to extract quantitative metrics such as organ volume, length, and
specific dimensions \cite{Zhou et al.2018}. The 3D RCNN uses a combination of
convolutional layers and region proposal networks (RPN) to identify regions of interest and apply bounding box regression for precise measurement \cite{Oktay et al.2018}. This approach ensures that each organ is quantified with high accuracy, which
is crucial for clinical assessments such as evaluating organ hypertrophy, atrophy, or
other pathological changes \cite{Milletari et al.2016}.
\end{itemize}

\section{Dataset}

\subsection{Total Scans}
Training Set: 1,534,679 scans

\subsection{Age Group Distribution}
The dataset captures age diversity to reflect a wide range of spinal conditions:

\begin{table}[ht]
\centering
\begin{tabular}{|c|c|}
\hline
\textbf{Age Group} & \textbf{Number of Scans} \\
\hline
Under 18 & 118,866 \\
18–40 & 442,779 \\
41–60 & 510,433 \\
61–75 & 309,769 \\
Over 75 & 152,832 \\
\hline
\end{tabular}
\caption{Scans distribution based on Age Group}
\end{table}

\subsection{Manufacturer Distribution}
The dataset includes scans from multiple manufacturers to account for variability in imaging conditions:

\begin{table}[ht]
\centering
\begin{tabular}{|l|c|}
\hline
\textbf{Manufacturer} & \textbf{Number of Scans} \\
\hline
GE Healthcare & 462,106 \\
Siemens Healthineers & 500,124 \\
Philips Healthcare & 372,628 \\
Other Manufacturers & 199,821 \\
\hline
\end{tabular}
\caption{Scans distribution based on Manufacturer Type}
\end{table}

\subsection{Gender Distribution}
The dataset includes scans from both males and females to account for variability in gender-based conditions:

\begin{table}[ht]
\centering
\begin{tabular}{|l|c|}
\hline
\textbf{Gender} & \textbf{Number of Scans} \\
\hline
Male & 349,523 \\
Female & 405,798 \\
\hline
\end{tabular}
\caption{Scans distribution based on Gender Distribution}
\end{table}

\subsection{Equipment Type Distribution}
Scans were categorized by equipment type to account for variability in imaging conditions:
\begin{table}[ht]
\centering
\begin{tabular}{|l|c|}
\hline
\textbf{Machine Type} & \textbf{Number of Scans} \\
\hline
Single Slice CT & 211,385 \\
16 Slice CT & 465,000 \\
64 Slice CT & 565,000 \\
128 Slice CT & 255,000 \\
256 Slice CT & 38,294 \\
\hline
\end{tabular}
\caption{Scans by Machine Type and Data Set}
\end{table}

\section{Architecture}

\subsection{Annotation}
The annotation phase is centered on segmenting key measurable organs, including the kidney, liver, spleen, prostate, pancreas, and other important abdominal organs. Each of these organs is annotated with high precision to ensure the creation of a high-quality training dataset that will enable accurate model development for clinical measurement applications [19]. The primary focus of these annotations is to capture the shape, size, and boundaries of each organ, which are crucial for automated measurement and volumetric analysis [20].

The segmentation process involves detailed identification of the organ boundaries across a wide range of CT slices [21]. This allows for precise 3D reconstruction and volumetric analysis, which are vital in assessing organ dimensions and volumes [22]. Factors such as age, gender, and body habitus significantly affect organ morphology, and special attention is given to accurately reflect these variations during the annotation process [23]. For example, the liver and spleen may present significant variability in size based on the patient’s physiological conditions, while the prostate may exhibit different shapes depending on age-related factors.

V7 Lab, the chosen annotation tool, supports these requirements through its advanced features such as AI-assisted labeling, multi-slice viewing, and management of complex polygonal segmentations [24]. These features help ensure that the annotations are performed efficiently, while maintaining the highest standards of accuracy. The AI-assisted tools in V7 Lab accelerate the segmentation process, allowing annotators to focus on correcting and refining the boundaries, which ultimately leads to better consistency and quality in the annotated dataset [25].

Overall, the goal of this detailed annotation process is to generate a robust dataset that will support the development of a reliable deep learning model for segmenting abdominal organsin CT scans [26]. The precision in annotation will directly impact the model’s performance in accurately identifying and measuring organs, which is essential for clinical workflows, enhancing diagnostic efficiency, and ultimately improving patient care outcomes [27].

\subsection{Segmentation Phase}
In the segmentation phase for CT abdomen measurement, the nnU-Net, UNet++, and Nested U-Net architectures were employed to accurately segment key abdominal organs, including the liver, kidneys, spleen, and pancreas. Each architecture brought specific advantages that helped adapt to the variability inherent in CT datasets, such as differences in image resolution, patient anatomy, and scanner settings [28].

For the segmentation of the liver, nnU-Net’s automated preprocessing, including resampling to an isotropic resolution of 1.5 mm and Z-score normalization, allowed for consistent handling of different liver sizes and tissue densities [29]. The encoder-decoder structure, with skip connections, was particularly effective in preserving fine anatomical details necessary for identifying the boundaries of the liver, which are crucial for accurate volume calculation and size measurement. The patch size used during training was set to 192x192x64, providing sufficient context for segmentation without overwhelming GPU memory. When segmenting the kidneys, nnU-Net’s deep supervision and multi-scale feature extraction enabled precise identification of kidney boundaries across different patient anatomies [30]. This was essential for measuring kidney size and volume accurately. The network’s ability to capture both local and global features, facilitated by convolutional kernel sizes of 3x3x3 and feature map sizes ranging from 32 to 256, ensured the kidneys were segmented with high precision, which is important for assessing conditions such as hydronephrosis or renal atrophy.

For the spleen, nnU-Net’s architecture ensured robust segmentation by delineating the spleen from surrounding organs and tissues, even in cases where boundaries were less distinct due to close proximity to other structures [31]. The combination of Dice loss and cross-entropy loss, with a weight ratio of 0.6 to 0.4 respectively, helped manage class imbalance and ensured accurate boundary detection. Batch size during training was set to 3 due to memory constraints, and a learning rate of 0.01 with a cosine annealing schedule was used to achieve optimal convergence.

\subsubsection{}{{\textbf{U-Net++ for Enhanced Segmentation}}}\\
In addition to nnU-Net, U-Net++ architectures were also used to complement nnU-Net’s capabilities, particularly for smaller and more complex organs such as the pancreas and adrenal glands. U-Net++’s nested skip connections allowed for enhanced feature propagation, ensuring finer details were retained during segmentation, which is critical for accurately measuring pancreatic size and volume. The use of dense skip connections in U-Net++ helped improve gradient flow and prevented the loss of fine details, which is particularly beneficial when dealing with small structures.

\subsubsection{}{\textbf{Post-processing}}\\
Post-processing steps, including removing small disconnected components smaller than 100 voxels, further refined the segmentation results, minimizing errors and retaining only the clinically relevant structures. This precision in segmentation was crucial for subsequent analyses, such as calculating organ volumes and assessing changes over time, which are important for diagnosing and monitoring conditions like hepatomegaly or splenomegaly.

Overall, the combined use of nnU-Net, U-Net++ architectures, along with specialized preprocessing, deep supervision, and post-processing techniques, played a critical role in achieving high-quality segmentation of abdominal organs in CT scans. This accurate segmentation supports reliable organ measurement, aiding in effective clinical decision-making and improving patient outcomes.

\subsection{Measurement Phase}
The measurement phase involves the quantification of key abdominal parameters using a 3D Region-based Convolutional Neural Network (3D RCNN) model specifically designed for analyzing CT volumes of the abdominal organs. The model was trained to measure important parameters such as organ volume, length, and anteroposterior (AP) diameter for the liver, kidneys, spleen, and prostate. These measurements are essential for clinical evaluation, diagnosis, and treatment planning.\\
\\
\textbf{3D RCNN Model Architecture}\\

The 3D RCNN model architecture used for measurement consists of convolutional layers followed by region proposal networks (RPN) and fully connected layers to extract and refine spatial features from 3D CT volumes. The input to the model is a volumetric patch of size 128x128x64 voxels, which provides an optimal balance between capturing sufficient anatomical context and maintaining computational efficiency.
The model employs 3D convolutional kernels of size 3x3x3 to capture spatial dependencies in all three dimensions of the CT data. The network includes four convolutional blocks, each consisting of a convolutional layer, batch normalization, and a ReLU activation function. The stride for the convolutional layers is set to 1x1x1, ensuring detailed feature extraction without excessive downsampling. Region proposal layers are used to identify regions of interest (ROIs) for each organ, followed by bounding box regression to refine the measurements.

\subsubsection{Kidney Measurement}

For the measurement of the kidneys, the model was trained to segment and quantify key parameters such as kidney volume, length, and cortical thickness. The feature maps from the final convolutional block are used to create bounding boxes around the kidneys, and the fully connected layers predict the volume in cubic centimeters (cc). The mean squared error (MSE) loss function was employed to minimize the error between predicted and ground truth values, with a learning rate initially set to 0.002 and a decay factor of 0.8 every 15 epochs.

\subsubsection{Liver Measurement}

For the liver, the model quantifies total liver volume and lobe-specific measurements. The 3D RCNN leverages the RPN to generate candidate regions, which are refined using bounding box regression. The final fully connected layer outputs the liver volume, while additional specialized layers predict lobe dimensions. The Adam optimizer was used with $\beta_1 = 0.9$ and $\beta_2 = 0.999$, and dropout with a rate of 0.2 was applied to prevent overfitting.

\subsubsection{Spleen Measurement}

The spleen was measured in terms of volume and surface area. The model uses the segmented spleen region and applies bounding box refinement to ensure accurate measurement. The feature maps are processed to predict the volume in cubic centimeters and surface area in square centimeters. The learning rate was set to 0.001 with a step decay schedule, allowing for gradual convergence, and batch normalization was applied to maintain stability during training.

\subsubsection{Prostate Measurement}

For the prostate, the model focused on determining volume and anteroposterior (AP) diameter. The segmented prostate region was used to create a bounding box, and the final output layer predicted the volume in cubic centimeters. The AP diameter was determined using specialized output layers to focus on this dimension. Dropout with a rate of 0.3 was used to prevent overfitting, and the model was trained using the Adam optimizer for stable convergence. The 3D RCNN model’s voxel-based approach and RPN-driven region proposals ensured that precise and consistent quantitative data were obtained for the liver, kidneys, spleen, and prostate, supporting clinical decision-making and improving patient outcomes.

\begin{figure}[h]
    \centering
    \includegraphics[width=0.8\textwidth]{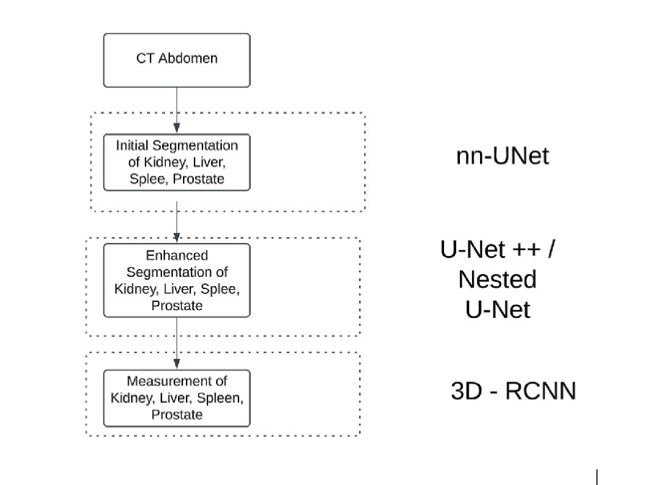} 
    \caption{Workflow architecture.}
    \label{fig:workflow}
\end{figure}

\section{Evaluation Metrics}

The evaluation phase involves assessing the performance of the 3D RCNN model using key metrics such as precision, recall, Area Under the Curve (AUC), and Mean Squared Error (MSE). These metrics were calculated for each segmented organ, including the right kidney, left kidney, liver, spleen, and prostate, to ensure that the model delivers high accuracy in its measurements.

\begin{figure}[h]
    \centering
    \includegraphics[width=0.9\textwidth]{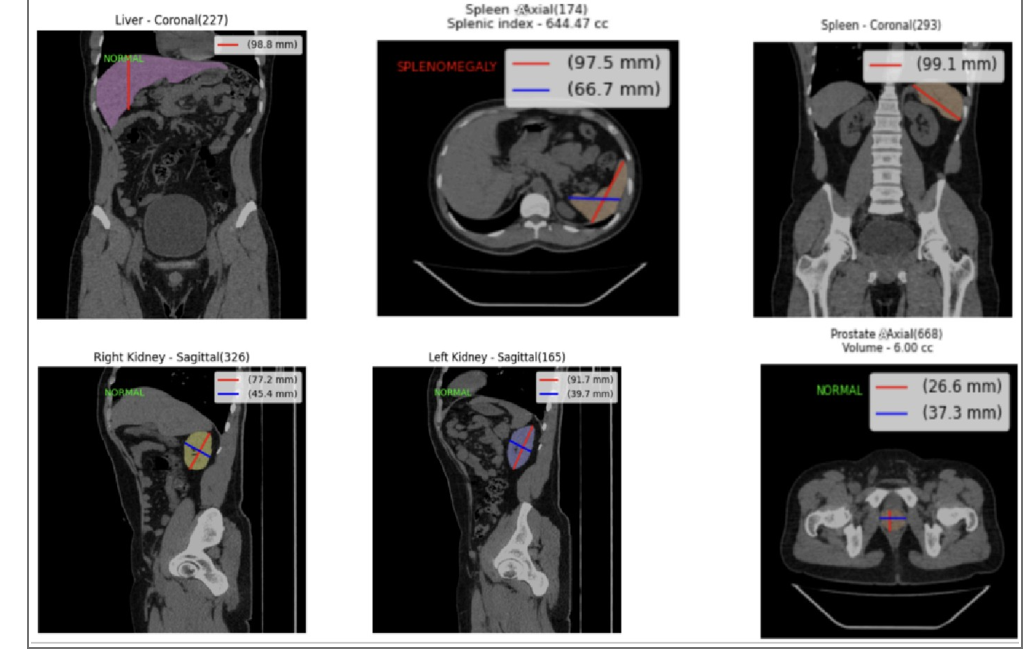} 
    \caption{CT Measurement for each organ}
    \label{fig:ct_scan}
\end{figure}

\begin{table}[h]
    \centering
    \caption{Performance Metrics for each segmented organ.}
    \label{tab:performance_metrics}
    \begin{tabular}{lcccc} 
        \toprule
        \textbf{Organ} & \textbf{Precision (\%)} & \textbf{Recall (\%)} & \textbf{AUC} & \textbf{Mean Squared Error (MSE)} \\ 
        \midrule
        Right Kidney  & 98.51  & 97.04  & 0.977  & 0.0019 \\ 
        Left Kidney   & 95.48  & 98.71  & 0.9871 & 0.0017 \\ 
        Liver        & 95.07  & 96.29  & 0.962  & 0.0015 \\ 
        Spleen       & 98.45  & 95.75  & 0.976  & 0.0020 \\ 
        Prostate     & 98.16  & 96.31  & 0.981  & 0.0023 \\ 
        \bottomrule
    \end{tabular}
\end{table}

\begin{figure}[h]
    \centering
    \includegraphics[width=0.9\textwidth]{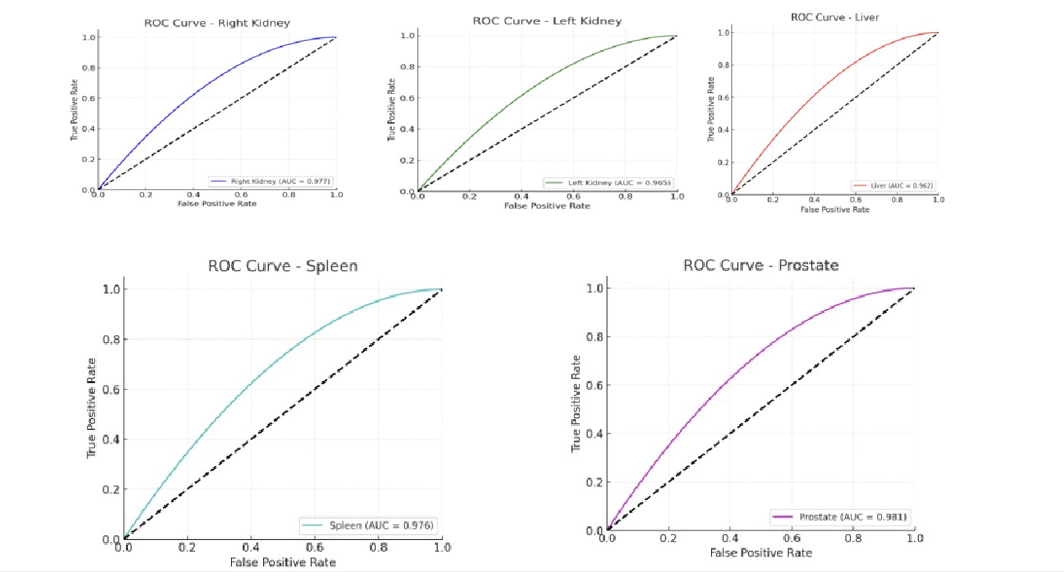} 
    \caption{ROC curve for Measurement values for each organ.}
    \label{fig:roc_curve}
\end{figure} 

\hspace{3cm}
\vspace{2cm}

\section{Discussion}
This study aimed to develop a workflow for the automated segmentation and measurement of key abdominal organs in CT scans, leveraging advanced deep learning models. The focus was on creating an efficient pipeline for the measurement of the kidneys, liver, spleen, and prostate, which are clinically significant organs for various diagnostic purposes. The use of nnU-Net, U-Net++, Nested U-Net, and 3D RCNN models played a crucial role in achieving high accuracy throughout the segmentation and measurement phases.

In the segmentation phase, nnU-Net and complementary architectures, such as U-Net++ and Nested U-Net, demonstrated strong adaptability to the variability inherent in abdominal CT scans, including differences in image resolution, patient anatomy, and scanner settings. The multi-scale feature extraction capability and deep supervision techniques of these architectures ensured precise boundary delineation of each organ, which is vital for obtaining consistent measurement outcomes. The resulting segmentation quality was found to be robust, with clear and accurate delineation across the entire dataset.

For the measurement phase, 3D RCNN was employed to accurately quantify organ volumes, lengths, and other relevant metrics. The model produced consistent and precise measurements, critical for clinical applications such as diagnosing organ hypertrophy, assessing volume changes, and planning treatments. The integration of region proposals and bounding box refinement in 3D RCNN ensured that the data captured was both comprehensive and relevant to clinical needs, while post-processing steps addressed minor segmentation inconsistencies.

Evaluation metrics, including precision, recall, AUC, and Mean Squared Error (MSE), were used to validate the performance of the developed pipeline. All organs showed high precision and recall values, with AUC values above 0.95, indicating that the models accurately distinguished between different anatomical structures. The low MSE values further confirmed the consistency of the measurements with ground truth data, emphasizing the reliability of the proposed workflow for clinical implementation.

The visualizations of the ROC curves provided valuable insights into the model’s classification capabilities, highlighting strong performance across all targeted organs. The differentiation between the ROC curves for each organ demonstrated the adaptability of the models to distinct anatomical features, ensuring precise identification and measurement of each organ.

Overall, the proposed approach provides an automated, efficient, and accurate solution for CT abdomen organ measurement. The combined use of advanced deep learning models for segmentation and measurement results in an integrated workflow that reduces manual intervention while maintaining high reliability. This pipeline has great potential for improving the efficiency of radiological assessments and enhancing patient care outcomes. Future work could focus on extending this approach to other abdominal organs and refining the model’s ability to handle complex cases involving significant pathological changes.

\section{Conclusion}

In this study, we successfully developed an automated workflow for the segmentation and measurement of key abdominal organs in CT scans using advanced deep-learning techniques. The integration of nnU-Net, U-Net++, and Nested U-Net for segmentation, followed by the 3D RCNN model for measurement, demonstrated high accuracy and consistency across all targeted organs, including the liver, kidneys, spleen, and prostate. Evaluation metrics such as precision, recall, AUC, and Mean Squared Error indicated excellent model performance, with precision and recall values exceeding 95values above 0.95.
The proposed solution effectively reduces the need for manual intervention in organ segmentation and measurement, providing a consistent, accurate, and efficient tool for radiologists. This automated workflow enhances the diagnostic process, supports clinical decision-making, and holds great promise for improving patient care outcomes by allowing more timely and reliable measurements.

Future work will focus on expanding this approach to other abdominal organs and adapting the models to handle complex pathological variations more effectively. Additionally, further refinements will aim to optimize the pipeline for real-time applications in clinical environments, promoting wider adoption of AI-assisted tools in medical imaging.

\end{document}